\documentstyle[11pt]{article}
\def\setfonts{%
\font\ssfbig=cmss10 scaled\magstephalf
\font\ssfscr=cmss8 %scaled\magstephalf
\font\ssfscrscr=cmss8
\newfam\ssffam
\textfont\ssffam=\ssfbig
\scriptfont\ssffam=\ssfscr
\scriptscriptfont\ssffam=\ssfscrscr
\def\ssf{\fam\ssffam}
}

\makeatletter
\@addtoreset{equation}{section}
\newdimen\normalarrayskip
\newdimen\minarrayskip
\normalarrayskip\baselineskip
\minarrayskip\jot
\newif\ifold \oldtrue \def\new{\oldfalse}
\def\arraymode{\ifold\relax\else\displaystyle\fi}

\def\@arrayskip{\ifold\baselineskip\z@\lineskip\z@
  \else
  \baselineskip\minarrayskip\lineskip2\minarrayskip\fi}
\def\@arrayclassz{\ifcase \@lastchclass \@acolampacol \or
\@ampacol \or \or \or \@addamp \or
 \@acolampacol \or \@firstampfalse \@acol \fi
\edef\@preamble{\@preamble
 \ifcase \@chnum
  \hfil$\relax\arraymode\@sharp$\hfil
  \or $\relax\arraymode\@sharp$\hfil
  \or \hfil$\relax\arraymode\@sharp$\fi}}
\def\@array[#1]#2{\setbox\@arstrutbox=\hbox{\vrule
  height\arraystretch \ht\strutbox
  depth\arraystretch \dp\strutbox
  width\z@}\@mkpream{#2}\edef\@preamble{\halign \noexpand\@halignto
\bgroup \tabskip\z@ \@arstrut \@preamble \tabskip\z@ \cr}%
\let\@startpbox\@@startpbox \let\@endpbox\@@endpbox
 \if #1t\vtop \else \if#1b\vbox \else \vcenter \fi\fi
 \bgroup \let\par\relax
 \let\@sharp##\let\protect\relax
 \@arrayskip\@preamble}
\@addtoreset{equation}{section}
\makeatother

\def\theequation{\thesection.\arabic{equation}}
\setfonts

\def\lvm{\leavevmode\hbox to\parindent{\hfill}}
\def\req#1{(\ref{#1})}

\def\BE{\begin{equation}}
\def\EE{\end{equation} }
\def\BA{\begin{array}}
\def\EA{\end{array}}

\def\bar{\overline}
\def\frac#1#2{{\textstyle{{#1}\over{#2}}}}
\def\ket#1{\bigl|{#1}\bigr\rangle}

\def\d{\partial}

\def\N#1{N\!=\!#1}
\def\SL#1{s\ell(#1)}
\def\SSL#1#2{s\ell(#1|#2)}
\def\OSP#1#2{osp(#1|#2)}

\def\half{{\textstyle{1\over2}}}
\def\threehalves{{\textstyle{3\over2}}}
\def\fourth{{\textstyle{1\over4}}}

\def\cB{{\cal B}}
\def\cC{{\cal C}}

\def\cE{{\cal E}}
\def\cF{{\cal F}}
\def\cG{{\cal G}}
\def\cH{{\cal H}}

\def\cL{{\cal L}}

\def\cQ{{\cal Q}}

\def\cT{{\cal T}}

\def\tensor{\otimes}

\def\NPB{Nucl.\ Phys.\ B}
\def\PRD{Phys.\ Rev.\ D}
\def\PLB{Phys.\ Lett.\ B}
\def\MPLA{Mod.\ Phys.\ Lett.\ A}

\def\IJMPA{Int.\ J.\ Mod.\ Phys.\ A}

\def\aP{\alpha_+}
\def\aM{\alpha_-}
\def\tbeta{\widetilde\beta}
\def\tgamma{\widetilde\gamma}
\def\barpsi{\bar\psi}
\def\Dbarphi{\d\bar\phi}
\def\Dphi{\d\phi}

\def\tBETA{{\widetilde\BETA}}
\def\tGAMMA{{\widetilde\GAMMA}}
\def\BETA{{\ssf B}}  %\EE
\def\GAMMA{{\Gamma}}

\def\Hplus{H^+}
\def\Hminus{H^-}
\def\Tm{T_{\rm m}}
\def\Gm{G_{\rm m}}
\def\barGm{\bar G_{\rm m}}
\def\Hm{H_{\rm m}}
\def\cm{c_{\rm m}}
\def\Tgr{\ssf T}
\def\DF{\d F}
\def\DU{\d U}
\def\Df{\d\varphi}
\def\Dv{\d v_*}
\def\DPHI{\d\Phi}
\def\DV{\d V_*}

\def\Iplus{I^+}
\def\iplus{I^{+{1\over2}}_{\phantom{H}}}
\def\Inaught{I^0}
\def\iminus{I^{-{1\over2}}_{\phantom{H}}}
\def\Iminus{I^-}

\def\Ipm{I^\pm}
\def\ipm{I^{\pm{1\over2}}_{\phantom{H}}}
\def\imp{I^{\mp{1\over2}}_{\phantom{H}}}

\def\ctop{{\ssf c}}
\def\qtop{{\ssf q}}
\def\AP{{\ssf a}_+}
\def\AM{{\ssf a}_-}
\def\qt{{\ssf q}}
\def\cTm{{\cal T}_{\rm m}}
\def\cGm{{\cal G}_{\rm m}}
\def\cQm{{\cal Q}_{\rm m}}
\def\cHm{{\cal H}_{\rm m}}

\def\Htplus{\cH^+}
\def\Htminus{\cH^-}
\def\BETAt{{\ssf B_{\rm t}}}
\def\GAMMAt{\Gamma_{\rm t}}
\def\Bt{B_{\rm t}}
\def\Ct{C_{\rm t}}

\def\BARPSI#1{{\bar\Psi}^{(#1)}_{{}_{\rm W}}} %\EE
\def\PSI#1{{\Psi}^{(#1)}_{{}_{\rm W}}}
\def\betaW{\beta_{{}_{\rm W}}}  %\EE
\def\gammaW{\gamma_{{}_{\rm W}}}
\def\DphiW{\d\varphi_{{}_{\rm W}}}
\def\DbarphiW{\d{\bar\varphi}_{{}_{\rm W}}}
\def\phiW{\varphi_{{}_{\rm W}}}
\def\barphiW{{\bar\varphi}_{{}_{\rm W}}} %\EE
\def\DcF{\d{\cal F}}
\def\DX{\d X}
\def\DbarX{\d\bar X}
\def\DcX{\d{\cal X}}

\def\emt{en\-ergy-mom\-ent\-um tensor}

\begin{document}
\addtolength{\baselineskip}{1pt}
\hfuzz=1pt
\begin{flushright}
{\tt hep-th/9604105}\\
\end{flushright}
\thispagestyle{empty}

\begin{center}
{\Large{\sc The Non-Critical $N=2$ String is an $\SSL21$ Theory}}\\[8pt]
{\large A.~M.~Semikhatov}\\[5pt]
{\small\sl I.~E.~Tamm Theory Division,
P.~N.~Lebedev Physics Institute, Russian Academy of Sciences}\\[10pt]
{\bf Abstract}\\[5pt]
\small Any non-critical $\N2$ string is shown to have an affine $\SSL21$
worldsheet symmetry in the {\it conformal\/} gauge.
\end{center}

\section{Introduction}\lvm
$\N2$ strings \cite{[Ade]} have been studied now for about two decades (see
\cite{[Marcus]} for a review).  The $(2,2)$ theory was first quantized a la
Polyakov in \cite{[FT]}, where the critical dimension was found to be 4. In
the BRST approach, this result was obtained and further discussed in
\cite{[DaL],[MM]}. That the target space is {\it complex two\/} dimensional
was realized in \cite{[OV23]}, where the geometry for the $\N2$ strings
was investigated and shown to be self-dual. The moduli and BRST cohomology of
the $\N2$ strings were studied in \cite{[Lech]}.  Recently, a very important
role was proposed for the heterotic $(2,1)$ strings in the
M-theory~\cite{[KM]}.

Non-critical string theories involve, in addition to the corresponding
(super)matter theory and ghosts, the (super)Liouville sector whose central
charge adds up with the matter central charge to the critical value.  The
$\N2$ string is described on the worldsheet by $\N2$ matter coupled to two
dimensional $\N2$ supergravity~\cite{[BS]}.  This string theory is singled
out for a number of reasons, one of them being that the bosonic ($\N0$) and
$\N1$ strings can be embedded into it~\cite{[BV],[FoF],[OP],[BOP],[BOh]}. On
the other hand, the $\N0,1,2$ non-critical string theories in the conformal
gauge have hidden $N+2$ twisted supersymmetry algebras on the
worldsheet~\cite{[GS2],[BLNW],[BLLS]}. The realizations of the $N+2$ algebras
provided by the non-critical string theories can be obtained via hamiltonian
reduction from the affine Lie superalgebras \cite{[BLNW],[IK]}.  This gives
rise to the proposal \cite{[BLLS],[RSS],[LLS]} to classify string theories in
terms of Ka\v c--Moody (super)algebras, the idea not unrelated to the
embeddings of different string theories into each other~\cite{[FoF]}.

The various relations between non-critical string theories and WZW models
have been studied from different points of view for a long time, and the
$N+2$ {\it supersymmetry\/} algebras are in fact a step towards the WZW
theories.  The simplest case corresponding to the bosonic string is the
affine $\SL2$ algebra; the $\N2$ superconformal algebra, which is realized in
the bosonic string, does indeed have many features in common with $\SL2$
(thus, the two algebras `share' the parafermions and have intimately related
singular vectors and in fact similar representation
theories~\cite{[ST23],[FST]}). Moreover, the realization of the $\N2$
superconformal algebra in the bosonic string carries over to an $\SL2$
representation in terms of an arbitrary matter theory tensored with several
free-field theories \cite{[S-sing]}.
This pattern extends to the $\N1$ superstring and the $\OSP12$ Ka\v c--Moody
algebra \cite{[HS]}.  Thus, in the two lowest cases ($\N0$ and $\N1$), by
tensoring the respective non-critical string with an additional free scalar,
one can reconstruct the ambient space of the hamiltonian reduction:  the
corresponding Ka\v c--Moody currents and the `Drinfeld--Sokolov' ghosts
(those that are needed in order to build the BRST complex for the hamiltonian
reduction).

In this paper we show that for the non-chiral $\N2$ (i.e., $(2,2)$) string
this reconstruction of, in that case, the $\SSL21$ Ka\v c--Moody superalgebra
can always be achieved directly in the space of fields of the non-critical
$\N2$ string in the conformal gauge, irrespective of the realization of its
matter sector.  Therefore, any non-critical $\N2$ string bears a
representation of the $\SSL21$ current algebra. In fact, at the level of
comparing the \emt s, the non-critical $\N2$ string is the same as the
$\SSL21$ current algebra together with two fermionic and two bosonic
first-order systems, the `Drinfeld--Sokolov' ghosts. Thus, while the
hamiltonian reduction of $\SSL21$ gives rise to the $N=2$ superconformal
algebra~\cite{[BO],[BLNW],[IK]}, dressing that `matter' theory with the $\N2$
gravity multiplet allows us to `invert' the reduction, i.e. to reconstruct
both the $\SSL21$ currents and the Drinfeld--Sokolov ghosts.

In section 2, we give the main construction of this paper, the $\SSL21$
currents realized in any non-critical $\N2$ string. Restricting this
construction to the $\OSP12$ and $\SL2$ subalgebras will result in a new
pattern of mappings between $\N2,1,0$ non-critical strings, which we consider
in section~3. An adaptation of the new realization of $\SSL21$ for the
topological string is given in section~4.  In section~5 we consider the
representation spaces and construct the $\SSL21$ highest-weight states in the
tensor product of certain $\N2$ matter Verma modules with free field modules,
and outline the corresponding KPZ-like formulation (which appears to have
become tautological by virtue of our construction of $\SSL21$).  In section~6
we consider a `non-stringy'$\!$, more economical, reformulation of the new
realization of $\SSL21$ in terms of $\N2$ matter dressed with free fields,
the latter no longer being those of the $\N2$ string; relation of the new
realization to the Wakimoto representations of $\SSL21$ is also considered
here.  Section~7 contains several concluding remarks, and the Appendix
summarizes our conventions.

\section{$\SSL21$ in the $\N2$ supergravity}\lvm In this section we will
present our main construction, a representation of the affine $\SSL21$
algebra \req{sl21opes} in terms of the fields of the $\N2$ noncritical
string. The $N=2$ matter is described by the \emt\ $\Tm(z)$, the
$U(1)$ current $\Hm(z)$ and two supersymmetry currents $\Gm(z)$ and
$\barGm(z)$. The operator products are written out in the Appendix. The $\N2$
matter central charge is
\BE
\cm = 3 - 6 q^2=-3 -6 k\,,
\EE
where the matter background charge $q^2$ satisfies\,\footnote{These formulae
simply introduce a new parameter, $k$; later this will become the level of
the $\SSL21$ algebra we are going to construct. The matter central charge
will thus be unambiguously related to the level.} $q^2 = k+1$.

In the non-critical $\N2$ string one also has the reparametrization ghosts
$bc$ and the $U(1)$ ghosts $\eta\xi$:
\BE
b(z)\,c(w) = {1\over z-w}\,,\qquad\eta(z)\,\xi(w) = {1\over z-w}\,,
\EE
and the bosonic ghosts
\BE
\tbeta(z)\,\tgamma(w) = {-1\over z-w}\,,\qquad
\beta(z)\,\gamma(w) = {-1\over z-w}\,,
\EE
and, finally, a super-Liouville sector represented in components by a free
complex scalar and a complex fermion:
\BE
\Dphi(z)\,\Dbarphi(w) = {-1\over(z-w)^2}\,,\qquad
\psi(z)\,\barpsi(w) = {1\over z-w}\,.
\label{Lopes}\EE

The energy-momentum tensor reads
\BE\new\BA{rcl}
\Tgr &=& \Tm - \d b  c - 2 b  \d c - \eta \d\xi
- \half \d\beta \gamma - \threehalves \beta \d\gamma
- \half \d\tbeta \tgamma - \threehalves \tbeta \d\tgamma\\{}&{}&{}
- \Dphi \Dbarphi + \half\, \d\Dbarphi + \half q^2\, \d\Dphi
+\half \d\psi \barpsi - \half \psi \d\barpsi
\EA\label{Tgr}\EE

We do not make any assumptions on the nature of the $\N2$ matter represented
by the \emt\ $\Tm$, the $U(1)$ current $\Hm$ and the supercurrents $\Gm$ and
$\barGm$. This can be any such theory, hence our construction is similar to
those of refs.~\cite{[GS2],[BLNW],[S-sing],[BLLS]} in that the matter sector
is only required to satisfy the appropriate (super)conformal algebra.  Another
remark is purely technical: the OPEs \req{Lopes} are invariant under
rescalings
\BE
\phi\mapsto a \phi\,,\qquad\bar\phi\mapsto a^{-1}\bar\phi\,,
\label{Lrescale}\EE
therefore only the product $\bar Q Q$ of the Liouville background charges
introduced into the \emt\ as $+ \half\,\bar Q\d\Dbarphi + \half Q\d\Dphi$
has an invariant meaning.  The central charge in the super-Liouville sector
is $3+6\bar Q Q=3+6q^2$.

Now we are going to use the fields $\Tm$, $\Hm$, $\barGm$, $\Gm$, $b,\,c$,
$\eta,\,\xi$, $\tbeta,\,\tgamma$, $\beta,\,\gamma$, $\Dphi,\,\Dbarphi$,
$\psi,\,\barpsi$ to construct the $\SSL21$ currents $E_1$, $E_2$, $E_{12}$,
$\Hminus$, $\Hplus$, $F_1$, $F_2$, $F_{12}$ that would satisfy the OPEs
\req{sl21opes}.

The upper-triangular currents are an easy part:
\BE\new\BA{rcl}
E_1 &=& \psi\, e^{\phi}\,,\\
E_2 &=& \barpsi\, e^{\phi}\,,\\
E_{12} &=& e^{2\phi}\,.
\EA\label{EEE}\EE
The choice of $\psi$ vs. $\barpsi$ is merely a convention; also, the
exponents in \req{EEE} may be changed by a common factor due
to~\req{Lrescale}.

Now we expect the matter ingredients $\Tm$, $\Gm$, $\barGm$, and $\Hm$ to
enter the lower-diagonal currents, i.e., the matter fields should go to where
they are to be found after the hamiltonian reduction. We thus expect $\Tm$ to
go into $F_{12}$. For $\barGm$ and $\Gm$ we also have natural candidates:
$F_1$ and $F_2$ respectively or vice versa (the choice is conventional in
view of the automorphism of the algebra \req{N2matter} that includes the
mapping $\barGm\leftrightarrow \Gm$). Further, we have $\Hplus$ to
incorporate the matter $\Hm$ current. Then, $\Hminus$ is a linear combination
of the remaining currents that one has in the $\N2$ non-critical string:
\BE\new\BA{rcl}
\Hminus &=&-\half \Dbarphi + \half(k - 3) \Dphi + b\,c +
\half \beta\,\gamma +   \half \tbeta\,\tgamma + \eta\,\xi\\
\Hplus&=&-\half \Hm + \half \psi\,\barpsi
\EA\label{Hminus}\EE
These currents are mutually orthogonal, normalized properly and satisfy the
required OPEs with the upper triangular currents \req{EEE}\,\footnote{We have
chosen the `real' basis for the two $\beta\gamma$ systems rather than the
`complex' basis in which $\beta^\pm(z)\,\gamma^\mp(w)=-1/(z-w)$.}.

The precise form of the rest of the algebra is less obvious, considering an
enormous number of {\it potentially\/} allowed terms. However, these actually
reduce to a reasonable amount; thus, $F_1$ is essentially proportional to
$\barpsi$, plus a piece that contains one of the matter supercurrents, as
expected:
\BE
F_1 =
\Bigl( \Gm - b c \,\barpsi\, -
\half \beta \gamma \,\barpsi -
\half \tbeta \tgamma \,\barpsi +
\half \Dbarphi \,\barpsi
+ \half (3 + k)  \Dphi \,\barpsi  -
\eta \xi \barpsi\,  -
\half  \Hm \barpsi -
(k+\half)  \d \barpsi
\Bigr)e^{-\phi}
\label{F1}\EE
Note that there is a symmetry
\BE
\barGm\to a\barGm\,,\quad\Gm\to a^{-1}\Gm
\label{N2scale}\EE
of the $\N2$ algebra, which allows us to introduce an arbitrary non-zero
parameter in front of $\Gm$ in \req{F1} (and then, accordingly, in front of
$\barGm$ in the subsequent formulae).

In \req{F1}, the notation is somewhat abused in the following manner. {\sl
Throughout the paper, the nested normal orderings are assumed from right to
left, as \ ${:}A\;{:}BC{:}\;{:}$}. However, in order to make the formulae
shorter, we write the vertex operators $\exp(a\phi)$ as common factors.  Thus
eq.~\req{F1} and similar formulae below should be understood by multiplying
every term in the parentheses with $e^{-\phi}$ and introducing the normal
ordering as explained.

Next, not surprisingly, $F_2$ is essentially proportional to $\psi$, plus a
term with the other matter supercurrent:
\BE
F_2 =
\Bigl(
(k+1)\,\barGm
+ b c\,  \psi  +
\half \beta  \gamma\, \psi +
\half \tbeta  \tgamma\,  \psi -
\half \Dbarphi\,  \psi
-\half (3 + k)  \Dphi\,  \psi  +
\eta  \xi\,  \psi  -  \half \Hm\, \psi   +
(k+\half)  \d \psi
\Bigr)e^{-\phi}
\label{F2}\EE
Now $F_1$ and $F_2$ give rise to $F_{12}$:
\BE\new\BA{rcl}
F_{12} &=&
\Bigl(
(k+1)  \Tm
- (k+1) \barGm \barpsi
+  \fourth \Hm \Hm
+ \half \Hm \psi \barpsi
- b c \beta \gamma  -
b c \tbeta \tgamma  +
b c \Dbarphi +
(3 + k)  b c \Dphi \\{}&{}&{}-
2  b c \eta \xi  -
k  b \d c  -
\fourth \beta \beta \gamma \gamma+
\half \beta \gamma \Dbarphi  +
(\frac{3}{2} + \frac{k}{2})  \beta \gamma \Dphi -
(\fourth + \frac{k}{2})  \beta \d \gamma  -
\fourth \tbeta \tbeta \tgamma \tgamma -
\half \tbeta \tgamma \beta \gamma \\{}&{}&{}+
\half \tbeta \tgamma \Dbarphi  +
(\frac{3}{2} + \frac{k}{2})  \tbeta \tgamma \Dphi -
(\fourth + \frac{k}{2})  \tbeta \d \tgamma  -
\fourth \Dbarphi \Dbarphi -
(\frac{3}{2} + \frac{k}{2})  \Dphi \Dbarphi -
\fourth (3 + k)^2  \Dphi \Dphi -
\eta \xi \beta \gamma \\{}&{}&{}-
\eta \xi \tbeta \tgamma  +
\eta \xi \Dbarphi +
(3 + k)  \eta \xi \Dphi  -
k  \eta \d \xi +
\Gm \psi +
(\fourth + \frac{k}{2})\psi \d \barpsi   -
(2 + k)  \d b c  -
(\frac{3}{4} + \frac{k}{2}) \d \beta \gamma \\{}&{}&{}-
(\frac{3}{4} + \frac{k}{2}) \d \tbeta \tgamma  +
\half (k+1) \d \Dbarphi  +
\half (3 + 4 k + k^2)  \d \Dphi -
(2 + k)  \d \eta \xi  -
(\frac{1}{4} + \frac{k}{2}) \d \psi \barpsi
\Bigr)e^{-2\phi}
\EA\label{F12}\EE
{\sl All the $\SSL21$ OPEs \req{sl21opes} are now satisfied\/} (and so are
those vanishing ones which are not shown in \req{sl21opes} explicitly).

\medskip

Since the $\N2$ matter is to be recovered as the result of the hamiltonian
reduction of $\SSL21$ \cite{[BO]}, the above construction can be viewed as an
`inversion' of the hamiltonian reduction. It is thus achieved by dressing the
$\N2$ matter into the $\N2$ string. A natural question is, how much bigger
the $\N2$ non-critical string is than the $\SSL21$ algebra just constructed,
that is, what is the `codimension' of $\SSL21$ in the $\N2$ string? It turns
out that the directions in the $\N2$ string field space orthogonal to the
$\SSL21$ algebra are naturally organized into precisely those ghost fields
that have to be introduced when performing the hamiltonian reduction
of $\SSL21$ (including the auxiliary fermionic system). These are two
fermionic $bc$ systems and two bosonic $\beta\gamma$ ones; we will call them
the {\sl Drinfeld--Sokolov ghosts\/}, in order to distinguish them from the
$\N2$ string ghosts.

The Drinfeld--Sokolov ghosts\ can be constructed along with the $\SSL21$
currents out of the same ingredients, the $\N2$ non-critical string, as
\BE\new\BA{rclcrclcl}
\tBETA  &=& \tbeta\, e^{-\phi}\,,&
\tGAMMA &=& \tgamma\,e^{\phi}\,,&\Rightarrow&\tBETA(z)\tGAMMA(w)=
{-1\over z-w}\\
\BETA  &=& \beta\, e^{-\phi}\,,&
\GAMMA &=& \gamma\, e^{\phi}\,,&\Rightarrow&\BETA(z)\GAMMA(w)={-1\over z-w}\\
B &=& b\, e^{-2\phi}\,,&
C &=& c\, e^{2\phi}\,,&\Rightarrow&B(z)C(w)={1\over z-w}\\
\cC &=& \xi\, e^{2\phi}\,,&
\cB &=& \eta\, e^{-2\phi}\,,&\Rightarrow&\cC(z)\cB(w)={1\over z-w}
\EA\label{DSghosts}\EE
Then we have the following fundamental identity, which can be interpreted as a
`completeness relation':
\BE
{\hbox{$\displaystyle T_{\rm Sug} + T_{\rm DS} =  \Tgr$}\strut}\,,
\label{identity}\EE
where $\Tgr$ is the $\N2$ string \emt\ \req{Tgr}, $T_{\rm Sug}$ is the
$\SSL21$ Sugawara \emt\ \req{N2Sug}, and the Drinfeld--Sokolov ghost \emt\ is
\BE
T_{\rm DS} = - \tBETA\, \d\tGAMMA - \BETA\,\d\GAMMA - B\,\d C - \cC\,\d\cB\,.
\EE

The identity \req{identity} tells us that the non-critical $N=2$ string is
the same, at the level of \emt s, as the $\SSL21$ superalgebra plus the
Drinfeld--Sokolov ghosts (we do not consider in this paper the subtle
questions of the definitions of the respective spaces of {\it physical\/}
states and relations between them).
Let us note that $\lim_{k\to-1}\,T_{\rm Sug}=\Tm + \ldots$ is finite in the
realization \req{EEE}--\req{F12}, and in fact the identity \req{identity}
remains valid as~$k\to-1$.

\pagebreak[3]

In the rest of the paper we consider several immediate consequences of the
construction \req{EEE}--\req{DSghosts}.

\medskip

A more compact expression for the $\SSL21$ currents can be obtained as
follows.  We introduce a scalar
\BE
\DF = -\half\Dbarphi - \half(3 + k)\Dphi +
b\,c + \half\beta\,\gamma + \half\tbeta\,\tgamma + \eta\,\xi
\label{DF}\EE
Then the $\SSL21$ currents rewrite as
\BE\new\BA{rcl}
\Hminus &=& \DF + k \Dphi\,,\\
F_1 &=& \Bigl(-\barpsi\,\DF  -
\half\Hm\,\barpsi + \Gm\,-
(k+\half)\d\barpsi\Bigr)e^{-\phi}\,,\\
F_2 &=& \Bigl(\psi\,\DF -
\half\Hm\,\psi + (k+1)\barGm +
(k+\half)\d\psi\Bigr)e^{-\phi}\,,\\
F_{12} &=& \Bigl(-\DF\,\DF - (k+1) \d\DF +
\Gm\,\psi + \fourth\Hm\,\Hm +
\half\Hm\, \psi\,\barpsi +
(\fourth + \frac{k}{2})\psi\,\d\barpsi \\{}&{}&{}-
(k+1)\barGm\,\barpsi + (k+1)\Tm +
(-\fourth - \frac{k}{2})\d\psi\, \barpsi\Bigr)e^{-2\phi}
\EA\label{compact}\EE
Let us add one more word of caution: as we have remarked, the exponential
factors in the above formulae are always assumed to multiply every monomial
in the parentheses and the monomials are normal ordered as,
e.g.,~${:}\DF\;{:}\DF\,e^{-2\phi}{:}\;{:}$.

It is now obvious, in particular, that the bosonic ghosts $\beta$, $\gamma$,
$\tbeta$, and $\tgamma$ enter the $\SSL21$ currents only through their
respective ghost currents \ ${:}\beta\,\gamma{:}$ and ${:}\tbeta\,\tgamma{:}$
(clearly, ${:}\beta\;{:}\gamma\,e^{-\phi}{:}\;{:}=
{:}\;{:}\beta\,\gamma{:}\;e^{-\phi}{:}$\,).

\section{Subalgebra embeddings and `projections' of non-critical strings}\lvm
In this section we show how the embeddings of subalgebras,
$\SL2\hookrightarrow\OSP12\hookrightarrow\SSL21$, can be used to derive
representations similar to \req{EEE}--\req{F12} for the $\SL2$ and $\OSP12$
current algebras. These representations will be in terms of the corresponding
($\N0$ and $\N1$) matter theories and a number of free fields.  All the free
fields originate from the $\N2$ string in the conformal gauge, and upon
reduction to $\N1$ and $\N0$ they will give rise to the respective ($\N1$ or
$\N0$) non-critical string ghosts and Liouville, plus an additional scalar;
the $\N2$ matter sector will mix with the $\N2$ Liouville superpartners to
produce the lower-supersymmetric matter theories. Thus we are going to
consider the following diagram:
\BE\BA{llccccc}
\parbox{80pt}{\small affine algebras}\qquad\qquad&&
  \SL2&\longleftarrow&\OSP12&\longleftarrow&\SSL21\\
{}&&\bigm\uparrow&{}&\bigm\uparrow&{}&\bigm\uparrow\\
\parbox{84pt}{\small non-critical strings, optionally with an\hfill\break
auxiliary scalar}&&
  (\N0)\tensor(\DV)&\longleftarrow&(\N1)\tensor(\Dv)&\longleftarrow&\N2\\
{}&&\bigm\downarrow&\swarrow&\bigm\downarrow&\swarrow&\bigm\downarrow\\
\parbox{80pt}{\small matter theories,\hfill\break
optionally with Liouville fermion(s)}&&
  (\N0)_{\rm m}&\longleftarrow&(\N1)_{\rm m}\tensor(\rho)&
\longleftarrow&(\N2)_{\rm m}\tensor(\barpsi,\psi)\\
{}&&\bigm\downarrow&\swarrow&\bigm\downarrow&\swarrow&\bigm\downarrow\\
\parbox{80pt}{\small matter theories}&&
  (\N0)_{\rm m}&{}&(\N1)_{\rm m}&{}&(\N2)_{\rm m}
\EA\label{square}\EE

We have adopted here the following `anti-mathematical' convention as regards
the direction of arrows. Consider two algebras $A$ and $B$ with generators
$(a_i)$ and $(b_j)$ respectively. Then constructing the {\it generators\/} of
$A$ in terms of those of $B$, $a_i=f_i(b_j)$, determines what is normally
denoted as the mapping $A\to B$, since it tells us that any $a_i$ is mapped
precisely into the respective $f_i$, viewed as an element of $B$.  Then we
would have to call, e.g., the mappings of ref.~\cite{[BV]} {\it
projections\/} rather than embeddings of string theories, like
$(\N2)\to(\N1)$.  We do not dare to challenge the established terminology
however, and will write $(\N1)\to(\N2)$ in the case when the $\N2$ fields are
constructed out of those of an $\N1$ theory.  {\it Then\/} the new mappings
we are going to discuss are directed as shown in~\req{square}. Thus, $A\to B$
reads `take the generators of $A$ and construct generators of $B$'$\!$.

Let us begin with the embedding $\OSP12\hookrightarrow\SSL21$: the $\OSP12$
currents can be identified~as
\BE\new\BA{rclrcl}
\Iplus &=& E_{12} \,,&\Iminus &=& F_{12}\,,\\
\iplus &=& E_1 + E_2\,,&\iminus &=& F_1 - F_2\,,\\
\Inaught &=& \Hminus
\EA\label{2osp}\EE
They satisfy the $\OSP12$ OPEs with precisely the above level $k$ (our
conventions for $\OSP12$ are specified in the Appendix).  By substituting
into \req{2osp} the above expressions for the $\SSL21$ currents, we will of
course get a realization of the $\OSP12$ algebra.  However that realization
will not be `irreducible' in the sense that the currents \req{2osp} depend
actually on only certain combinations of the $\N2$ string fields. These
special combinations turn out to make up an $\N1$ non-critical string theory
onto which we can `{\it project\/}' from the $\N2$ string. Let us describe
therefore how this is carried out. We are going to introduce a new basis in
the space of fields of the $\N2$ string in such a way that some of the new
fields will be the $\N1$ string fields, while the others will decouple from
the string.

First, we realize the $\N2$ Liouville fermion in terms of two real fermions
$\lambda$ and $\rho$,
\BE
\rho(z)\,\rho(w) = -{1\over z-w}\,,\qquad
\lambda(z)\,\lambda(w) = {1\over z-w}\,,
\label{rholambda}\EE
as
\BE
\psi = \frac{1}{\sqrt{2}}(\rho + \lambda)\,,\qquad
\barpsi =\frac{1}{\sqrt{2}}(\lambda - \rho)\,,
\label{psirule}\EE
which, clearly, is an invertible transformation.  Then, as we are going to
see, $\rho$ will be a part of the $\N1$ super-Liouville, while $\lambda$ will
participate in constructing an `effective' $\N1$ matter theory:  {\sl For any
$\N2$ matter with a non-critical central charge $\cm=-3-6k\neq6$, there
is a mapping $(\N2)_{\rm m}\tensor(\lambda)\to(\N1)_{\rm m}$ to an $\N1$
theory\/}, which is defined by
\BE\new\BA{rcl}
G^{{\phantom{y}}}_{N=1} &=&
-\sqrt{2}\aP\Gm + \frac{1}{\sqrt{2}}(\aM + \aP)\barGm +
\aP\Hm\, \lambda\,,\\
T^{{\phantom{y}}}_{N=1} &=& (1 - \aP^2)\Tm + \sqrt{2}\aP^2\Gm\, \lambda +
\half\aP^2\Hm\, \Hm +
\frac{1}{\sqrt{2}}(\aP^2-1)\barGm\, \lambda +
(\aP^2-\half)\d\lambda\, \lambda\,,
\EA\label{matterembed}\EE
\BE
\aP=-\frac{1}{\aM}\,,\qquad \aM^2=2k+3\,.
\EE
The $\N1$ superconformal algebra can be checked for the generators thus
constructed:
\BE\new\BA{rcl}
T^{{\phantom{y}}}_{N=1}(z) T^{{\phantom{y}}}_{N=1}(w) &=&
{d^{{\phantom{y}}}_{N=1}/2 \over(z-w)^4}
    + {2 T^{{\phantom{y}}}_{N=1}(w)\over(z-w)^2} +
{\d T^{{\phantom{y}}}_{N=1}(w)\over z-w}\,,\\
T^{{\phantom{y}}}_{N=1}(z) G^{{\phantom{y}}}_{N=1}(w) &=&
{3/2\,G^{{\phantom{y}}}_{N=1}(w)\over(z-w)^2}
    + {\d G^{{\phantom{y}}}_{N=1}(w)\over z-w} \,,\\
G^{{\phantom{y}}}_{N=1}(z) G^{{\phantom{y}}}_{N=1}(w) &=&
{2/3\,d^{{\phantom{y}}}_{N=1}\over(z-w)^3}
    + {2 T^{{\phantom{y}}}_{N=1}(w)\over z-w}\,,
\EA\label{matter1}\EE
the central charge being
\BE
d^{{\phantom{y}}}_{N=1} = \frac{15}{2} - 3\aM^2 - 3 \aP^2\,
\label{d1}\EE
by virtue of eqs.~\req{N2matter},\,\req{rholambda}.  We call the theory
($G^{{\phantom{y}}}_{N=1}$, $T^{{\phantom{y}}}_{N=1}$),
eqs.~\req{matterembed}, the `effective' $\N1$ matter.

The remaining fields of the $\N2$ string are $\Dbarphi$, $\Dphi$, $b$, $c$,
$\beta$, $\gamma$, $\tbeta$, $\tgamma$, and $\xi$, $\eta$. We now change the
basis in the space of these free fields; as the new independent fields we
take:  {\sl i)\/}~the ghosts $\beta$, $\gamma$, $b$, $c$, \ {\sl ii)\/}~{\it
the Drinfeld--Sokolov ghosts\/} $\tBETA$, $\tGAMMA$, $\cC$, $\cB$, and {\sl
iii)\/}~two real bosons $\Dv$ and $\Df$, with
\BE\Dv(z)\,\Dv(w) = {1\over(z-w)^2}\,,\qquad
\Df(z)\,\Df(w) = -{1\over(z-w)^2}\,.
\EE
Thus, in addition to \req{psirule}, we express the fields $\Dbarphi$,
$\Dphi$, $\tbeta$, $\tgamma$, $\eta$, $\xi$ as
\BE\new\BA{rclrcl}
\Dphi &=& \aP(\Df - \Dv)\,,&
\Dbarphi &=& 2 \cB\cC + \tBETA\,\tGAMMA -
  (\half\aM - \frac{3}{2}\aP)(\Df+\Dv)\,,\\
\tbeta &=& \tBETA\, e^{\aP(\varphi-v_*)}
\,,&
\tgamma &=& \tGAMMA\,e^{-\aP(\varphi-v_*)}
\,,\\
\eta &=& \cB\,e^{2\aP(\varphi-v_*)}
\,,&
\xi &=& \cC\,e^{-2\aP(\varphi-v_*)}
\EA\label{rules}\EE
The last two lines are nothing but the first and the last lines of
\req{DSghosts} respectively, however with $\tBETA$ and $\tGAMMA$ (bosons), and
$\cC$ and $\cB$ (fermions) now viewed as independent fields with the operator
products \ $\tBETA(z)\tGAMMA(w)=-1/(z-w)$, $\cB(z)\cC(w)=1/(z-w)$.  The free
fermion $\rho$, which is what remains of the $\N2$ Liouville superpartners
after $\lambda$ has gone to construct the $\N1$ matter, does not mix with
other fields.

In the new basis, the currents \req{2osp} rewrite as
\BE\new\BA{rcl}
\Iplus&=&e^{2\aP(\varphi - v_*)}\,,\\
\iplus &=&\sqrt{2}\rho\,e^{\aP(\varphi - v_*)}\,,\\
\Inaught &=&-\frac{3}{2}\aP \Df + \half(\aM + 3 \aP) \Dv + b\, c +
\half\beta\, \gamma\,,\\
\iminus &=&\Bigl(\sqrt{2} b\, c\, \rho\,  +
\frac{1}{\sqrt{2}}\beta\, \gamma\, \rho +
\frac{\aM}{\sqrt{2}}\Df\, \rho  +
\frac{\aM}{\sqrt{2}}G^{{\phantom{y}}}_{N=1}  +
\frac{1}{\sqrt{2}}(\aM^2 - 2 ) \d \rho\Bigr)e^{-\aP(\varphi - v_*)}\,,\\
\Iminus &=&\Bigl(-b\, c\, \beta\, \gamma\,  -
\aM b\, c\, \Df +
(\frac{3}{2} - \half \aM^2) b\, \d c  -
\fourth\beta\, \beta\, \gamma\,\gamma -
\half\aM\beta\, \gamma\,\Df  \\{}&{}&{}+
(\half - \fourth \aM^2) \beta\,\d \gamma
- \fourth\aM^2\Df\, \Df -
\half\aM\rho\, G^{{\phantom{y}}}_{N=1} +
\half \aM^2 T^{{\phantom{y}}}_{N=1} -
(\half + \half\aM^2) \d b\, c  \\{}&{}&{}-
\fourth\aM^2\d \beta\, \gamma +
\fourth (\aM - \aM^3) \d \Df +
(\fourth\aM^2 - \half) \d \rho\,\rho\, \Bigr)e^{-2\aP(\varphi-v_*)}
\EA\label{osp}\EE
and in this form represent an intrinsic-$\N1$ construction~\cite{[HS]}:
given {\it any\/}
$\N1$ matter theory ($G_{N=1}$, $T_{N=1}$) determined by
\req{matter1},\,\req{d1}, coupling it to the $\N1$ supergravity multiplet
$\Df$, $\rho$, $bc$, $\beta\gamma$, and additionally tensoring with an
auxiliary scalar $\Dv$, we would obtain a realization of the $\OSP12$ current
algebra, with the $\OSP12$ OPEs \req{osp12OPE} satisfied and the level being
$k={1\over2}(\alpha_-^2-3)$.

The other two $\N2$ Drinfeld--Sokolov ghost systems $BC$ and $\BETA\GAMMA$,
re-expressed in terms of the new fields introduced in \req{rules},
\BE\new\BA{rclcrcl}
\BETA &=& \beta\,e^{-\aP(\varphi-v_*)}\,,&
\GAMMA &=& \gamma\,e^{\aP(\varphi-v_*)}\,,\\
B &=& b\,e^{-2\aP(\varphi-v_*)}\,,&
C &=& c\,e^{2\aP(\varphi-v_*)}\,,
\EA\label{N1DS}\EE
remain the `Drinfeld--Sokolov' ghosts for the $\OSP12$ algebra:  the $\OSP12$
theory with the twisted Sugawara energy-momentum tensor, plus the \emt\ for
$BC$ and $\BETA\GAMMA$,
\BE\new\BA{rcl}
{\widetilde T}_{\OSP12}&=& 2 \aP^2 ( \Inaught\,\Inaught + \half\Iplus\,\Iminus+
\half\Iminus\,\Iplus + \fourth\iplus\,\iminus -
\fourth\iminus\,\iplus) + \d \Inaught\,,\\
T_{\rm GH}&=&{}-\d B\,C - 2B\,\d C - \half\d\BETA\,\GAMMA -
\frac{3}{2}\BETA\,\d\GAMMA
\EA\EE
rewrite precisely as the $\N1$ supergravity plus an extra $U(1)$ theory
represented by $\Dv$:
\begin{eqnarray}
{\widetilde T}_{\OSP12} + T_{\rm GH}&=&
T^{{\phantom{y}}}_{N=1}
- 2b\, \d c - \d b\, c- \frac{3}{2}\beta\,\d\gamma -
  \half\d \beta\,\gamma -
  \half\Df\,\Df   +
  \half(\aP - \aM)\d\Df - \half\d\rho\,\rho \nonumber
\\{}&{}&{}+ \half\Dv\,\Dv + \half(\aM - \aP)\d\Dv\,,
\label{emtN1}\end{eqnarray}
which is an $\OSP12$--($\N1$) analogue of \req{identity}.  The central charge
in the $\Df$-$\rho$ sector is the required ${15\over2} + 3\aM^2 + 3\aP^2$
(cf.~\req{d1}).

Eqs.~\req{osp},\,\req{N1DS} are the construction of the $\OSP12$ current
algebra from~\cite{[HS]}, which is valid for any non-critical $\N1$ string
tensored with an auxiliary scalar.  Now we see that this is also the $\OSP12$
representation induced from the $\SSL21$ algebra realized in the $\N2$
string. The subalgebra embedding $\OSP12\hookrightarrow\SSL21$ thus allows us
to map the respective non-critical string theories and, another level down,
the matter theories as in~\req{matterembed}, see~\req{square}.

To return to the interpretation of $(T^{{\phantom{y}}}_{N=1},
G^{{\phantom{y}}}_{N=1})$ as the `effective' $\N1$ matter made out of the
ingredients of the $\N2$ string, we see that the $bc$ and $\beta\gamma$
ghosts remain in their capacities of the reparametrization and the
`gravitino' ghosts respectively, while $\Df$ and $\rho$ are the components of
the $\N1$ super-Liouville (thus, as expected, just one of the $\N2$
supersymmetries gets broken). On the other hand, $\Dv$, $\tBETA\tGAMMA$ and
$\cB\cC$ decouple from the $\N1$ string.  It may be worth noting that the
$\cB\cC$ and $\tBETA\tGAMMA$ ghosts make up a {\it topological\/}
$bc\beta\gamma$ theory, so one may practice in proving the cohomological
triviality of modding out the $\cB\cC\tBETA\tGAMMA$ system.  The
identifications of the different sectors can be summarized as follows:
$$\new
\BA{l}
\beta\,\gamma\quad b\,c\quad\Tm\quad\Hm\quad\barGm\quad\Gm\quad\underbrace{
 \psi\quad\bar\psi}\quad\underbrace{\Dphi\quad\Dbarphi\quad\tbeta\,
\tgamma\quad\xi\,\eta}\\
\hbox{\rlap{$\displaystyle
\beta\,\gamma\quad b\,c\quad\Tm\quad\Hm\quad\barGm\quad\Gm\quad\overbrace{
\lambda\quad\rho}\quad\overbrace{\Df\quad\Dv\quad\tBETA\,\tGAMMA
\quad\cC\,\cB}$}}
\phantom{\beta\,\gamma\quad b\,c\quad}\underbrace{\phantom{\Tm\quad\Hm\quad
\barGm\quad\Gm\quad\lambda}}
\phantom{\quad\rho}
\\
\underbrace{\underbrace{\beta\,\gamma\quad b\,c~~\,\qquad
\overbrace{T_{N=1}\qquad G_{N=1}}
\quad~\,\quad\rho\quad\Df}_{N=1\ \rm string}\quad\Dv}
\quad\tBETA\,\tGAMMA\quad\cC\,\cB\\
\quad~\qquad\qquad\overbrace{\OSP12\qquad B\,C\qquad\BETA\,\GAMMA}
\qquad\qquad\qquad\;\tBETA\,\tGAMMA\quad\cC\,\cB
\EA
$$
We would like to stress that this picture reflects nothing but a simple fact
of the subalgebra embedding $\OSP12\hookrightarrow\SSL21$ of the $\OSP12$
algebra from the bottom line into the $\SSL21$ algebra associated with the
top line of the diagram.

\medskip

This is not yet the end of the story: one can reduce further down to $\N0$
theories, which would correspond to taking just the (common) $\SL2$
subalgebra of $\SSL21$ and $\OSP12$.  We thus start with the theory whose
\emt\ is given by the RHS of \req{emtN1}  with $\aP^2\neq-1$ ($k\neq-2$) and
further rearrange the fields with the aim to single out the non-critical
bosonic string.  First of all, the $\N1$ matter \req{matter1} tensored with
the $\rho$ fermion `projects' onto the $\N0$ matter:
\BE\new\BA{rcl}
T^{{\phantom{y}}}_{N=0} &=& {1\over1+ \aP^2}T^{{\phantom{y}}}_{N=1}
- {\aP\over1 + \aP^2}G^{{\phantom{y}}}_{N=1}\,\rho +
          {1 - 2 \aP^2\over2(1 + \aP^2)}\d \rho\,\rho\\
{}&=&{}
\frac{1}{k+2}\Bigl((k+1)\Tm + \Gm\,\psi +
  \fourth\Hm\,\Hm +
  \half \Hm\,\psi\,\barpsi  -
  (k+1)\barGm\,\barpsi \\{}&{}&{}+
  \fourth(1 + 2k)\psi\,\d\barpsi-
  \fourth(1 + 2 k)\d\psi\,\barpsi)\Bigr)
\EA\label{TN0}\EE
where we have also re-expressed $T^{{\phantom{y}}}_{N=0}$ through the fields
of the $\N2$ string\,\footnote{\label{foot:cc}Which shows that the throughout
`reduction' to the $\N0$ matter (and hence to the bosonic string) is possible
from the $\N2$ string whenever $\cm\neq9$.}. In the
following, however, we will not expand the $\N1$ fields in terms of the $\N2$
ingredients.  As is easy to check, the operator products \req{matter1},
\req{rholambda} alone allow one to show that $T^{{\phantom{y}}}_{N=0}$ \ {\it
is\/} an \emt\ with central charge
\BE
d^{{\phantom{y}}}_{N=0}=13-6(k+2)-\frac{6}{k+2}\,.
\label{N0cc}\EE

To `project' onto the whole $\N0$ non-critical string, we need to identify
the corresponding Liouville field. To this end we mix the free fields as
follows.  Recall that in the $\N1$ setting we have, in addition to the $\N1$
matter, the ghosts $bc$ and $\beta\gamma$, the Liouville system $\Df$ and
$\rho$, and the auxiliary scalar $\Dv$. As we saw in the first equation in
\req{TN0}, $\rho$ goes to construct the `effective' $\N0$ matter. Among the
remaining free fields, we choose a new basis:  as the new independent fields,
we take the bosonic $\N1$ Drinfeld--Sokolov ghosts $\BETA\GAMMA$, along with
two scalars $\DPHI$ and $\DV$ with the operator products
\BE\DV(z)\,\DV(w) = {1\over(z-w)^2}\,,\qquad
\DPHI(z)\,\DPHI(w) = -{1\over(z-w)^2}\,,
\EE
introduced via
\BE\new\BA{rcl}
\Df &=& \aP \BETA\GAMMA
+ \frac{1}{\sqrt{1+\aP^2}}(\DPHI + \aP^2\DV)\,,\\
\Dv &=& \aP \BETA \GAMMA + \sqrt{1 + \aP^2}\DV\,,\\
\beta&=& \BETA \exp\Bigl(\frac{\aP}{\sqrt{1+\aP^2}}(\Phi - V_*)\Bigr)\,,\quad
\gamma~{}={}~ \GAMMA \exp\Bigl(\frac{-\aP}{\sqrt{1+\aP^2}}(\Phi - V_*)\Bigr)
\EA\EE
Then the $\SL2$ currents rewrite as
\BE\new\BA{rcl}
\Iplus&=&
e^{{\sqrt{2}}a_+(\Phi-V_*)} \\
\Inaught&=&
-\sqrt{2}a_+\DPHI + \bigl(\frac{1}{\sqrt{2}}a_- +
     {\sqrt{2}}a_+\bigr) \DV  + b c \\
\Iminus&=&
\Bigl(
(2 - a_-^2)b \d c
     - {\sqrt{2}}a_-
      \DPHI b c
   - a_-^2\d b c
\\{}&{}&{}
     + a_-^2\bigl(T_{N=0}-\half\DPHI \DPHI   -
      \frac{1}{\sqrt{2}}( a_- + a_+ )
          \d\DPHI\bigr)\Bigr)
e^{-{\sqrt{2}}a_+(\Phi-V_*)}
\EA\label{K}\EE
where $a_-=-1/a_+$ and
\BE
a_+=\frac{\sqrt{2}\aP}{\sqrt{1 + \aP^2}}\,,
\EE
whence $a_-^2=k+2$. The $\SL2$ currents are thus expressed through the \emt\
$T_{N=0}$, the $bc$ (reparametrization) ghosts, the $\N0$ Liouville $\DPHI$
and the auxiliary scalar $\DV$.  In fact, any $\N0$ matter theory with
central charge \req{N0cc} can be dressed into an $\SL2$ algebra of level $k$
by tensoring with the ghosts, a `Liouville' scalar and an auxiliary scalar.
This construction, known from~\cite{[S-sing]}, is therefore recovered from
the embedding $\SL2\hookrightarrow\OSP12$ \cite{[HS]}.

The fermionic Drinfeld--Sokolov ghosts $BC$ that we had in the $\N1$ context,
remain the Drinfeld--Sokolov ghosts for the $\SL2$ algebra. They are
re-expressed in terms of the new ingredients as
\BE
B=b\,e^{-\sqrt{2} a_+(\Phi - V_*)}\,,\qquad
C = c\,e^{\sqrt{2} a_+(\Phi - V_*)}\,,
\EE
which allows us to show that the  (twisted) $\SL2$ Sugawara energy-momentum
tensor, plus the \emt\ for the $BC$ ghosts, evaluate as the $\N0$ string with
an auxiliary $\DV$ scalar:
\BE\new\BA{l}
\frac{1}{k+2} ( \Inaught\,\Inaught + \half\Iplus\,\Iminus +
\half\Iminus\,\Iplus) + \d\Inaught  - \d B\,C - 2 B\,\d C\\
\qquad{}=
T_{N=0} -
2b\,\d c - \d b\, c
- \half\DPHI\,\DPHI - \frac{1}{\sqrt{2}}(a_- - a_+)\d\DPHI\\
\qquad{}+ \half\DV\,\DV + \frac{1}{\sqrt{2}}(a_- - a_+)\d\DV\,.
\EA\EE
This is the $\N0$ version of the `completeness relation'.  The central charge
in the $\DPHI$ (`Liouville')
sector is \ $13+6(k+2)+{6\over k+2}$, as it should.

\medskip

A remark is in order concerning the mapping ($\N1$ matter,
$\rho)\to($`effective' $\N0$ matter), eq.~\req{TN0}, induced by the embedding
$s\ell(2)\hookrightarrow osp(1|2)$: it relates the central charges as
\BE
d_{N=1} =
{ d_{N=0}^2 + 13d_{N=0} - 5 \pm
  (d_{N=0} - 1)\sqrt{(25 - d_{N=0})(1 - d_{N=0})}
   \over 2(2 + d_{N=0})}
\EE
When $d_{N=0}\to-2$, the `$+$'-branch gives a finite value
$d_{N=1}=-{5\over2}$.  The inverse relation reads
\BE
d_{N=0} = {(d_{N=1} - {27\over2})(d_{N=1} + {1\over2}) \pm
      ({3\over2}-d_{N=1}) \sqrt{({27\over2}-d_{N=1})({3\over2}-d_{N=1})}
\over2d_{N=1} - 27}
\label{toefff}
\EE
The mapping \req{toefff} can be applied to any $N\!=\!1$ $(p,q)$ minimal
model tensored with a $\rho$ fermion. The $N\!=\!1$ central charge is then
mapped as
\BE
\frac{3}{2}-3\frac{p}{q}-3\frac{q}{p}\mapsto\left
\{\new\BA{l}13-6\frac{p+q}{2q}-6\frac{2q}{p+q}\\
13-6\frac{p+q}{2p}-6\frac{2p}{p+q}
\EA\right.\EE
depending on the sign in \req{toefff}, and the `effective' $\N0$ matter theory
thus gets identified with a model from the $N\!=\!0$ minimal series.

\medskip

The main feature that distinguishes the $\SSL21/\N2$ case is that,
in contrast to $\OSP12$ and $\SL2$, the construction of the $\SSL21$ currents
does {\it not\/} involve any extra free fields in addition to the fields of
the respective supergravity in the conformal gauge.  Note also that the
representation \req{K} of $\SL2$ in terms of the $\N0$ non-critical string
$(T_{N=0},\DPHI,b,c)$ tensored with the $\DV$ scalar blows up at $k\to-2$,
which naturally corresponds to the infinite matter central charge
$d_{N=0}\to\infty$; similarly, the construction \req{osp} of $\OSP12$ works
only for $k\neq-{3\over2}$, which means having finite $\N1$ matter central
charges $d_{N=1}\neq\infty$.  On the other hand, the realization
\req{EEE}--\req{F12} of the $\SSL21$ algebra remains valid for any finite
level $k$.

In the {\it critical\/} $\N2$ string, with $\cm=6$, $q^2=-\half$, the
Liouville sector decouples from the string, but not from the $\SSL21$
algebra, which continues smoothly to the corresponding level $k=-{3\over2}$.
The construction \req{EEE}--\req{F12} is therefore somewhat `less natural',
from the stringy point of view, for the critical $\N2$ string, since it
requires either introducing an `external'
$\d\phi\d\bar\phi\,\psi\bar\psi$-theory or structuring the critical matter
into the `proper matter' and the `Liouville' theory whose central charges add
up to the critical value.

\section{Topological strings}\lvm In this section we reformulate
eqs.~\req{EEE}--\req{F12} for the case of the {\it topological\/} strings
\cite{[VV]}. We thus start with the topological conformal matter
represented by the generators $\cTm$, $\cGm$, $\cQm$, and $\cHm$, which
satisfy the twisted-$\N2$ algebra
\BE\new\BA{rclcrcl}
\cTm(z)\cTm(w)&=&{2\cTm(w)\over(z-w)^2}+{\d\cTm\over z-w}\,,&\qquad&
\cHm(z)\cHm(w)&=&{\ctop/3\over(z-w)^2}\,,\\
\cTm(z)\cGm(w)&=&{2\cGm(w)\over(z-w)^2}+{\d\cGm\over z-w}\,,&{}&
\cTm(z)\cQm(w)&=&{\cQm(w)\over(z-w)^2}+{\d\cQm\over z-w}\,,\\
\cTm(z)\cHm(w)&=&\multicolumn{5}{l}{{\ctop/3\over(z-w)^3}+
{\cHm(w)\over(z-w)^2}+{\d\cHm\over z-w}\,,}\\
\cHm(z)\cGm(w)&=&{\cGm\over z-w}\,,&{}&\cHm(z)\cQm(w)&=&-{\cQm\over z-w}\,,\\
\cQm(z)\cGm(w)&=&\multicolumn{5}{l}{{\ctop/3\over(z-w)^3}
-{\cHm(w)\over(z-w)^2}+{\cTm\over z-w}\,,}
\EA\label{topOPE}\EE
with the topological central charge
\BE\ctop=-3-6k\,.\EE
We will not introduce more notations and will have $\bar\psi$ denote the
gradient of the `topological' fermion $\chi$, for which
$\d\chi(z)\psi(w)=1/(z-w)$. Similarly, we will preserve the notation
$\Dbarphi$ for what is often called $\d\pi$ in the topological
string~\cite{[VV]}. The ghost sector will consist of the bosonic and
fermionic ghosts $\beta$, $\gamma$ and $b$, $c$ \cite{[VV]}.  We thus have
the following fields making up the topological string, with their conformal
dimensions given in Table~1.
\begin{center}
\renewcommand{\arraystretch}{0}
\begin{tabular}{|c|c|c|c|c|c|c|c|c|c|c|c|}
\hline
\multicolumn{12}{|c|}{\rule{0pt}{1.5pt}}\\\hline
\strut$b$&$c$&$\beta$&$\gamma$&$\cTm$&$\cGm$&$\cQm$&$\cHm$&$\Dbarphi$&
$\Dphi$&$\psi$&${\barpsi}^{{\phantom{h}}}$\\
\hline
\strut$2$&$-1$&$2$&$-1$&$2$&$2$&$1$&$1$&$1$&$1$&$0$&$1$\\
\hline
\multicolumn{12}{|c|}{\rule{0pt}{1.5pt}}\\\hline
\end{tabular}\\[6pt]
{\small{\bf Table 1.} Conformal dimensions of the topological string
ingredients}
\end{center}

Then the formulae \req{Hminus}--\req{F12} can very easily\,\footnote{The
reason being that $F_1$, $F_2$ and $\Hminus$ (and hence $F_{12}$) depend only
on the combination $\beta\,\gamma + \tbeta\,\tgamma + 2\eta\,\xi + 2bc$, as
can be seen from \req{DF}--\req{compact}.} be modified to produce the
$\SSL21$ algebra constructed out of the fields from Table~1:
\BE\new\BA{rcl}
\Htplus&=&-\half \cHm + \half \psi\,\barpsi\,,\\
\Htminus &=&
-\half\Dbarphi + \half(-3 + k)\Dphi +
\half(\frac{9}{2\qt^2} + \frac{\qt^2}{3})b\,c +
\half(\frac{9}{2\qt^2} - \frac{\qt^2}{3})\beta\,\gamma\,,\\
\cF_1 &=&
\Bigl( \cGm - \half(\frac{9}{2\qtop^2} + \frac{\qtop^2}{3}) b c \,\barpsi -
  \half (\frac{9}{2\qtop^2} - \frac{\qtop^2}{3})\beta \gamma \,\barpsi +
  \half \Dbarphi \,\barpsi \\{}&{}&{}
+ \half (3 + k)  \Dphi \,\barpsi -
  \half  \cHm \barpsi -
  (k+\half)  \d \barpsi
\Bigr)e^{-\phi}\,,\\
\cF_2 &=&
\Bigl(
(k+1)\cQm +
\half(\frac{9}{2\qt^2} + \frac{\qt^2}{3})b\,c\,\psi +
  \half(\frac{9}{2\qt^2} - \frac{\qt^2}{3})\beta\,\gamma\,\psi -
  \half\Dbarphi\,\psi \\{}&{}&{}
- \half(3 + k)\Dphi\,\psi -
  \half\cHm\,\psi  +
  (k+\half)\d\psi
\Bigr)e^{-\phi}\,,
\EA\label{Ft}\EE
while $\cE_i=E_i$, eqs. \req{EEE} (and $\cF_{12}$ uniquely determined by
$\cF_1$ and $\cF_2$). Here $\qt$ is a free parameter, which will be related
to the topological
Liouville background charge. To see how this can be done, consider the
`topological' analogue of the `completeness relation' \req{identity}.  The
formulae \req{Ft} differ from \req{Hminus}--\req{F2} essentially by a
redefinition of the fields by means of an $O(2,2)$ rotation in the space of
the four ghost currents $bc$, $\beta\gamma$, $\tbeta\tgamma$, and $\eta\xi$:
the $bc$ and $\beta\gamma$ currents in \req{Ft} are particular linear
combinations of the four currents from \req{Hminus}--\req{F2}. Then the
construction of the `Drinfeld--Sokolov' ghosts changes accordingly: in the
topological setting, we define, instead of \req{DSghosts},
\BE\new\BA{rclrcl}
\BETAt  &=& \beta\, e^{-(9/(2 \qt^2) - \qt^2/3)\phi}\,,&
\GAMMAt &=& \gamma\, e^{(9/(2 \qt^2) - \qt^2/3)\phi}\,,\\
\Bt &=& b\, e^{-(9/(2\qt^2) + \qt^2/3)\phi}\,,&
\Ct &=& c\, e^{(9/(2\qt^2) + \qt^2/3)\phi}\,.
\EA\label{topDS}\EE
Now, we evaluate the \emt\ for these `Drinfeld--Sokolov'\,\footnote{In the
topological context, the name may be misleading, but we continue to use it in
order to keep track of the various ghost theories.} ghosts and add it with
the $\SSL21$ Sugawara \emt; the latter, however, has to be {\it doubly
twisted\/}, then it follows from the previous formulae that
\BE\new\BA{l}
\frac{1}{k+1}\Bigl(\Htminus\, \Htminus - \Htplus\,\Htplus +
\half \cE_{12}\, \cF_{12} + \half \cF_{12}\, \cE_{12} +
\half \cE_1\,\cF_1 - \half \cF_1\,\cE_1 -
\half \cE_2\,\cF_2 + \half \cF_2\,\cE_2\Bigr) + \d\Htminus + \d\Htplus\\
\qquad\qquad{} - \d\Bt\,\Ct - 2 \Bt\,\d\Ct -
\d\BETAt\,\GAMMAt - 2 \BETAt\,\d\GAMMAt\\
\qquad\qquad\qquad\qquad{}=
\cTm -\d b\,c - 2 b\,\d c - \d\beta\,\gamma - 2 \beta\,\d\gamma -
\Dbarphi\,\Dphi + \d\psi\,\barpsi + (\qt^2 + k) \d\Dphi\,,
\EA\label{topidentity}\EE
thereby recovering (for $\bar\psi=\d\chi$) the \emt\ of the topological
string.

Dimensions of the `composite' fields evaluated with respect to the \emt\ on
the LHS of \req{topidentity}, are as follows:
\begin{center}
\renewcommand{\arraystretch}{0}
\begin{tabular}{|c|c|c|c|c|c|c|c|c|c|c|c|}
\hline
\multicolumn{12}{|c|}{\rule{0pt}{1.5pt}}\\\hline
\strut$\cE_1$&$\cE_2$&$\cE_{12}$&$\Htplus$&$\Htminus$&$\cF_1$&$\cF_2$&
$\cF_{12}$&$\BETAt$&$\GAMMAt$&$\Bt$&${\Ct}^{{\phantom{h}}}$\\
\hline
\strut$0$&$1$&$0$&$1$&$1$&$2$&$1$&$2$&$2$&$-1$&$2$&$-1$\\
\hline
\multicolumn{12}{|c|}{\rule{0pt}{1.5pt}}\\\hline
\end{tabular}\\[6pt]
{\small{\bf Table~2.} Conformal dimensions of doubly-twisted $\SSL21$
currents and Drinfeld--Sokolov ghosts}
\end{center}

\section{Constructing the highest-weight states and a KPZ-like picture}\lvm
In this section, we go further with the analysis of the new representation
for the $\SSL21$ algebra and consider briefly how states in the
representation space of $\SSL21$ can be built in terms of the `stringy'
ingredients. We begin with the $\SSL21$ construction induced from the
topological string, since this would allow us to have fewer fields without
missing any general feature.

In the ghost sectors, we choose the $sl(2)$-invariant vacua~\cite{[FMS]}
\BE
b_{\geq-1}\ket{0}_{bc}=0\,,\quad c_{\geq2}\ket{0}_{bc}=0\,,\qquad
\beta_{\geq-1}\ket{0}_{\beta\gamma}=0\,,\quad
\gamma_{\geq2}\ket{0}_{\beta\gamma}=0\,.
\EE
In the $\barpsi\psi$-sector, similarly,
\BE
\barpsi_{\geq0}\ket{0}_{\barpsi\psi}=0\,,\quad
\psi_{\geq1}\ket{0}_{\barpsi\psi}=0\,.
\EE
Consider further the topological matter sector \req{topOPE}.  There are
two different types of highest-weight conditions for the $\N2$ algebra:
{\sl i)\/}~those that define `chiral' primary states~\cite{[LVW]},
\BE\new\BA{rcl}
\cQ_{\geq1}\,\ket{h,\ell}&=&\cG_{\geq0}\,\ket{h,\ell}~{}={}~
\cL_{\geq1}\,\ket{h,\ell}~{}={}~
\cH_{\geq1}\,\ket{h,\ell}~{}={}~0\,,\\
\cH_0\,\ket{h}&=&h\,\ket{h,\ell}\,,\\
\cL_0\,\ket{h}&=&\ell\,\ket{h,\ell}
\EA\label{masshw}\EE
(where $\cT(z)=\sum\cL_n\,z^{-n-2}$, $\cG(z)=\sum\cG_n\,z^{-n-2}$,
$\cQ(z)=\sum\cQ_n\,z^{-n-1}$, $\cH(z)=\sum\cH_n\,z^{-n-1}$), and {\sl
ii)\/}~those that define `chiral' and at the same time `BRST-invariant'
primary states,
\BE\new\BA{rcl}
\cQ_{\geq0}\,\ket{h}&=&\cG_{\geq0}\,\ket{h}~{}={}~
\cL_{\geq1}\,\ket{h}~{}={}~
\cH_{\geq1}\,\ket{h}~{}={}~0\,,\\
\cH_0\,\ket{h}&=&h\,\ket{h}\,,\\
\cL_0\,\ket{h}&=&0\,.
\EA\label{tophw}\EE
In \cite{[ST23]}, the latter were called {\it topological\/} primary
states, but the term may be a bit overloaded in the present paper.

The $\SSL21$ highest-weight states can be obtained by tensoring the
corresponding $\N2$ highest-weight states with the free-field vacua. The
parameters have to be related in order that the $\SSL21$ state satisfy the
required conditions.  In this paper we only consider how this works for the
$\N2$ topological highest-weight states.
In fact, the $\SSL21$ highest-weight states come out tensored with the
`Drinfeld--Sokolov' ghost vacua: consider
\BE
\ket{p,h}_{\SSL21}\tensor\ket{0}_{\Bt\Ct}\tensor\ket{0}_{\BETAt\GAMMAt}=
\ket{0}_{bc}\tensor\ket{0}_{\beta\gamma}\tensor
\ket{0}_{\barpsi\psi}\tensor\ket{h}\tensor\ket{p}_{\bar\phi\phi}
\label{ket}\EE
where $\ket{p}_{\bar\phi\phi}$ is a state corresponding to the vertex
operator $\exp p\,\phi$.  From \req{topDS} it follows that \req{ket} {\it
is\/} a proper vacuum for each of the `Drinfeld--Sokolov' ghost systems.  For
the $\SSL21$ currents~\req{Ft}, similarly, we find (with the mode expansions
of the currents performed in accordance with their dimensions from Table~2):
\BE\new\BA{rclcrclcrcl}
(\cE_1)_{\geq1}\ket{p,h}_{\SSL21}&=&0\,,&&
(\cE_2)_{\geq0}\ket{p,h}_{\SSL21}&=&0\,,&&
(\cE_{12})_{\geq1}\ket{p,h}_{\SSL21}&=&0\,,\\
(\cF_1)_{\geq0}\ket{p,h}_{\SSL21}&=&0\,,&&
(\cF_2)_{\geq1}\ket{p,h}_{\SSL21}&=&0\,,&&
(\cF_{12})_{\geq1}\ket{p,h}_{\SSL21}&=&0\,,\\
\Htminus_{\geq1}\ket{p,h}_{\SSL21}&=&0\,,&&
\Htplus_{\geq1}\ket{p,h}_{\SSL21}&=&0\,,
\EA\EE
In addition, we find the following eigenvalues of the Cartan operators:
\BE\new\BA{rcl}
\Htplus_0\ket{p,h}_{\SSL21}&=&-\frac{h}{2}\,\ket{p,h}_{\SSL21}\,,\\
\Htminus_0\ket{p,h}_{\SSL21}&=&\frac{p}{2}\,\ket{p,h}_{\SSL21}\,,
\EA\EE
which actually label the states we are considering\,\footnote{The third
parameter, the eigenvalue $k$ of the central element of the algebra, is
omitted from the notations.}.

Those modes of the currents that do not annihilate the state thus constructed,
do generate a module from $\ket{p,h}_{\SSL21}$. In terms of the `component'
fields from Table~1, these are evaluated e.g. as
\BE\new\BA{rcl}
(\cF_1)_{-1}\ket{p,h}_{\SSL21}&=&
\ket{0}_{bc}\tensor\ket{0}_{\beta\gamma}\tensor
\ket{0}_{\barpsi\psi}\tensor(\cGm)_{-1}\ket{h}\tensor\ket{p}_{\bar\phi\phi}\\
{}&{}&{}-\frac{h+p}{2}\,\ket{0}_{bc}\tensor\ket{0}_{\beta\gamma}\tensor
\barpsi_{-1}\ket{0}_{\barpsi\psi}\tensor\ket{h}\tensor\ket{p}_{\bar\phi\phi}
\,,\\
(\cF_2)_{0}\ket{p,h}_{\SSL21}&=&\frac{p-h}{2}\,
\ket{0}_{bc}\tensor\ket{0}_{\beta\gamma}\tensor
\psi_0\ket{0}_{\barpsi\psi}\tensor\ket{h}\tensor\ket{p}_{\bar\phi\phi}\,,
\EA\label{generated}\EE
etc.

However, the absence of the annihilation condition for $(\cF_{12})_{0}$
(which would be $(F_{12})_{1}$ in the standard setting with the currents
being of dimension~1, as corresponds to the {\it un\/}twisted Sugawara \emt)
means that we are in the $\SSL21$ module with infinitely many highest-weight
vectors (thus, in a sense `without' a highest-weight vector, assuming this to
be unique); our aim, on the other hand, is to build an $\SSL21$ Verma module
with a unique highest-weight state out of the ingredients of the $\N2$
string.  This requires introducing further relations between the parameters
involved, which entails the restriction to a class of `overdetermined'
$\SSL21$ highest-weight states. An obvious possibility is to set $p=h$, then
we see from \req{generated} that $(\cF_{2})_{0}$ would vanish on the
highest-weight state, and {\it hence\/} so would do~$(\cF_{12})_{0}$.

For the honest $\N2$ string, such a mechanism of having, in
that case, $(F_{2})_{0}$ and $(F_{12})_{1}$
vanish simultaneously on the highest-weight state works as
follows. First of all, the construction similar to \req{ket} follows by
tensoring both sides of \req{ket} with the missing ghosts and then performing
the appropriate $O(2,2)$ mixing in the ghost sector.  Then we would have
$(E_1)_{\geq0}\approx0,\, (E_2)_{\geq0}\approx0,\,(E_{12})_{\geq0}\approx0,\,
(F_1)_{\geq1}\approx0,\,(F_2)_{\geq1}\approx0,\, (F_{12})_{\geq2}\approx0$ on
the thus constructed state.  Further, the \emt\ \req{Tgr} endows it with
dimension $(p-h)/2$. Now, requiring this to vanish will also result in the
vanishing of the total $U(1)$ charge\,\footnote{To avoid confusion, note that
our normalization of the Liouville currents $\Dphi$ and $\Dbarphi$ differs
from that of ref.~\cite{[BLLS]} by a transformation of the form \req{Lopes}.}
and, on the other hand, {\it will also be the condition for
$(F_2)_0^{{\phantom{y}}}$ and $(F_{12})_1^{{\phantom{y}}}$ to annihilate the
highest-weight state\/} (with the other highest-weight conditions already
satisfied). Thus the requirement that the free fields dress the $\N2$
topological highest-weight state into a {\it massless\/} $\SSL21$ state
produces the restricted class of the $\SSL21$ highest-weight conditions.

Thus, the $\N2$ string appears to be a `self-KPZ' theory, for which the
appropriate generalizations of the DDK ansatz \cite{[Da],[DK]} in the
conformal gauge and the KPZ formulation \cite{[P-gr],[KPZ]} can be carried
out on the same field space.

\medskip

We would like to stress that one has the {\it Verma\/} module over the
highest-weight vector \req{tophw} in the $\N2$ matter sector (hence, in
particular, no `accidental' vanishing of singular vectors can occur unless
one bosonizes the matter sector through free fields, thereby going over to
Fock modules).

\section{A `non-stringy' reformulation and the Wakimoto representations
of the affine $\SSL21$}\lvm
Our construction of section~2 employed the full set of fields of the $\N2$
string
in the conformal gauge.  As a result, we were able to construct not only the
$\SSL21$ currents, but also the Drinfeld--Sokolov ghosts; the theory
$\SSL21\tensor({\rm DS\ ghosts})$ was singled out by the `completeness
relation' \req{identity}, whose RHS gives the $\N2$ string \emt. However,
when one is not particularly interested in the string, one may prefer having
fewer ingredients at the starting point and, accordingly, only the $\SSL21$
currents (with no extra `Drinfeld--Sokolov' ghosts) as the output. Such a
reformulation is easy to arrive at, by a change of basis in the space of the
free fields:  in addition to \req{DF}, one introduces
\BE
\DU = -\half\Dbarphi - \half(3 - k)\Dphi +
         b\,c + \half\beta\,\gamma + \half\tbeta\,\tgamma + \eta\,\xi
\EE
Then $\DF$ and $\DU$, with
\BE
\DF(z)\DF(w)={-k/2\over(z-w)^2}\,,\qquad
\DU(z)\DU(w)={k/2\over(z-w)^2}\,,
\label{DFDU}\EE
can be considered as new independent free fields, and the $\SSL21$ currents
at level $k\neq0$ rewrite in terms of $\Tm$, $\Hm$, $\Gm$, $\barGm$, $F$,
$U$, $\psi$, and $\bar\psi$ as
\BE\new\BA{rcl}
E_1 &=& \psi\, e^{{1\over k}(U-F)}\,,\qquad\qquad
E_2\;{}={}\,\barpsi\, e^{{1\over k}(U-F)}\,,\qquad
E_{12}\,{}={}\,e^{{2\over k}(U-F)}\,,\\
\Hplus&=&-\half \Hm + \half \psi\,\barpsi\,,\qquad
\Hminus = \DU\,,\\
F_1 &=& \bigl(\Gm - \barpsi\,\DF  -
\half\Hm\,\barpsi -
(k+\half)\d\barpsi\bigr)e^{-{1\over k}(U-F)}\,,\\
F_2 &=& \bigl((k+1)\barGm + \psi\,\DF -
\half\Hm\,\psi  +
(k+\half)\d\psi\bigr)e^{-{1\over k}(U-F)}\,,\\
F_{12} &=&
\bigl(-\DF\,\DF - (k+1) \d\DF + (k+2)T_{N=0}\bigr)e^{-{2\over k}(U-F)}\,,
\EA\label{eco}\EE
where the `effective' \emt\ $T_{N=0}$ stands for the second expression in
\req{TN0}. Equations~\req{eco} show, in particular, that $\DU$ is nothing but a
`bosonization' of the diagonal current from the $\SL2$
subalgebra\footnote{One should probably have normalized the $\DF$ and $\DU$
currents to $\pm1$ over the poles in \req{DFDU}, in which case the exponents
in \req{eco} would acquire the factors $\pm\sqrt{{1\over2k}}$,
$\pm\sqrt{{2\over k}}$, and the \emt\ in~\req{sugeco} would take the
canonical form. An essential point is that, in whatever normalization, the
currents $\DU$ and $\DF$ have opposite signatures.}; thus, factoring it away
would leave us with the appropriate super-parafermions, which, too, would be
represented in terms of an arbitrary $\N2$ matter with the appropriate
central charge and the free fields $F$, $\bar\psi$, $\psi$.  The Sugawara
\emt\ \req{N2Sug} evaluates in the realization \req{eco} as
\BE
T_{\rm Sug} = \Tm - \frac{1}{k}\DF\,\DF - \d\DF -
\half\psi\,\d\bar\psi + \half\d\psi\,\bar\psi + \frac{1}{k}\DU\,\DU
\label{sugeco}\EE
The principle that \req{eco} inherits from the construction of section~2 is
that the $\N2$ matter sector can be arbitrary.

\bigskip

Recently, the {\it Wakimoto\/} $\SSL21$ modules \cite{[BO]} and their
relation to the $\N2$ strings were considered in~\cite{[BKT]}. As was pointed
out in \cite{[S-sing]} on the $\SL2$ example, representations of the type
of those considered in the present paper (in particular, eqs.~\req{eco}) can
be mapped onto the Wakimoto representations at the expense of explicitly
bosonizing the matter sector through some free fields.  The same is true for
the $\SSL21$ algebra, however there are some new features due to the
existence of two simple root systems not related via a proper affine Weyl
group reflection.

The fact that $\SSL21$ admits two different simple root systems manifests
itself in the hamiltonian reduction: there exist two different ways to perform
the hamiltonian reduction of $\SSL21$ to $\N2$ matter~\cite{[BLNW]}.  Another
consequence of the existence of two simple root systems has been worked out
in~\cite{[BKT]}, where a `second' Wakimoto representation for $\SSL21$ was
found, related to the appropriate `second' Gauss decomposition.  An
interesting question thus is whether the two essentially different Wakimoto
representations are related in any reasonable way to the
realization~\req{eco}. As we have noted, the $\SL2$ experience tells us that
the Wakimoto ingredients can indeed be constructed as soon as one explicitly
`bosonizes' the matter sector in terms of free fields.  Such a bosonization
is just what we have been carefully avoiding, since an important property of
the realization \req{eco} is that it is valid for an {\it arbitrary\/} matter
sector. Now, in contrast to our strategy so far, we do bosonize the $\N2$
matter sector $(\Tm,\,\Gm,\,\Hm,\,\barGm)$ and see if the necessary
ingredients of the Wakimoto representation can be constructed in the
resulting space of free fields.

However, there are {\it two\/} essentially different ways to `bosonize' the
$\N2$ matter.  The first one is through a complex scalar and a fermion,
\BE
\DbarX(z)\,\DX(w) = {1\over(z-w)^2}\,,\qquad
\chi(z)\,\bar\chi(w) = -{1\over z-w}\,,
\label{Xchi}\EE
as in \cite{[MSS],[OS],[Ito]} (modulo simple field redefinitions):
\BE\new\BA{rcl}
\Tm &=& \half\chi\, \d\bar\chi -
   \half\d\chi\,\bar\chi + \DbarX\,\DX - \half\d\DbarX - \half(k+1) \d\DX\,,\\
\Hm &=& \DbarX - (k+1)\DX - \chi\,\bar\chi\,,\\
\barGm &=& -\bar\chi\,\DbarX + (k+1)\d\bar\chi\,,\\
\Gm &=& \chi\,\DX - \d\chi\,.
\EA\label{Xmatter}\EE
The operator products \req{Xchi} are invariant under rescalings similar
to~\req{Lrescale}, and this freedom, both in the $\DbarX\DX$ and
$\bar\chi\chi$ sectors, has been used in order to minimize the number of
times $(k+1)^{-1}$ appears in~\req{Xmatter} and the subsequent formulae.

The following formulae show why we needed the fields $\chi$, $\bar\chi$,
$\DbarX$ and $\DX$ to represent the $\N2$ matter: the Wakimoto ingredients
are constructed in terms of the free fields from \req{eco} and \req{Xmatter} as
\BE\new\BA{rclcrcl}
\betaW &=& e^{{2\over k}(U-F)}\,,\\
\gammaW &=&\multicolumn{5}{l}{\Bigl(\DF + \half(k+1)\DX + \half\DbarX +
        \half\psi\,\barpsi - \half\chi\,\bar\chi -
        (k+1)\barpsi\,\bar\chi\Bigr)e^{-{2\over k}(U-F)} \,,}\\
\DbarphiW &=&\DbarX - \frac{k + 1}{k} \DU +
   \frac{k + 1}{k}\DF \,,&&
\DphiW &=&\DX + \frac{1}{k}\DF - \frac{1}{k}\DU\,,\\
\BARPSI1 &=& \chi\,e^{-{1\over k}(U-F)} + (k+1)\barpsi\,e^{-{1\over k}(U-F)}
\,,&&
\PSI1 &=& \bar\chi\,e^{{1\over k}(U-F)} \,,\\
\BARPSI2 &=& \barpsi\,e^{{1\over k}(U-F)} \,,&&
\PSI2 &=& \psi\,e^{-{1\over k}(U-F)} + (k+1)\bar\chi\,e^{-{1\over k}(U-F)}
\EA\label{fromwak}\EE
which would be impossible had we had only the generators $\Tm$, $\Gm$,
$\barGm$ and $\Hm$ (and the remaining fields $F$, $U$, and $\psi,\bar\psi$).
The Wakimoto fields \req{fromwak} can be considered as
an independent bosonic beta-gamma system, a complex boson and two fermionic
$bc$ systems with operator products
\BE\new\BA{rclrcl}
\betaW(z)\,\gammaW(w)&=&{-1\over z-w}\,,&
\DphiW(z)\,\DbarphiW(w)&=&{1\over(z-w)^2}\,,\\
\PSI1(z)\,\BARPSI1(w) &=& {-1\over z-w}\,,&
\PSI2(z)\,\BARPSI2(w) &=& {1\over z-w}\,.
\EA\label{wakope}\EE
Now we insert the fields \req{fromwak} into the Wakimoto realization for
$\SSL21$. A convenient form of the latter is obtained from the formulae of
refs.~\cite{[BO],[BKT]} by simple field redefinitions which are automorphisms
of the defining operator product expansions:
\BE\new\BA{rcl}
E_1 &=& \PSI2\,\betaW - (k+1)\PSI1 \,,\qquad E_2~{}={}~\BARPSI2\,,\qquad
E_{12}~{}={}~\betaW\,,\nonumber\\
\Hplus &=& -\half\DbarphiW + \half(k+1)\DphiW -
    \half\PSI1\,\BARPSI1 +  \half\PSI2\,\BARPSI2\,,\nonumber\\
\Hminus &=& -\half\DbarphiW - \half(k+1)\DphiW  -
    \half\PSI1\,\BARPSI1 -  \half\PSI2\,\BARPSI2 + \gammaW\,\betaW\,,\nonumber\\
F_1 &=& - \BARPSI2\,\gammaW + \DphiW\,\BARPSI1 -
         \d\BARPSI1\,,\nonumber\\
F_2 &=& \PSI2\,\gammaW\,\betaW - \DbarphiW\,\PSI2 -  (k+1)\PSI1\,\gammaW  -
          \PSI2\,\PSI1\,\BARPSI1 + k \d\PSI2\,,\nonumber\\
F_{12}&=&
-\gammaW\,\gammaW\,\betaW + \DbarphiW\,\gammaW + (k + 1)\DphiW\,\gammaW
 - k\d\gammaW
  + \PSI1\,\BARPSI1\,\gammaW
+ \PSI2\,\BARPSI2\,\gammaW \\{}&{}&{}+
\PSI2\,\d\BARPSI1 -\DphiW\,\PSI2\,\BARPSI1\,.\nonumber
\EA\label{wak}\EE
{\sl After the substitution \req{fromwak}, the currents \req{wak} become
the respective $\SSL21$ currents~\req{eco} with the $\N2$ matter generators
taken in the bosonized form~\req{Xmatter}\/}.

Furthermore, the standard Wakimoto screening operators
\cite{[BO],[IK],[BLNW]} map under \req{fromwak} into the $\N2$ {\it matter\/}
screenings \cite{[DvPZ],[YZ]}:
\BE\new\BA{rcl}
\PSI1\,e^{\phiW}&=&\bar\chi\, e^{X},\\
\Bigl(\BARPSI1\,\betaW -
   (k+1)\BARPSI2\Bigr)\,e^{{1\over k+1}\barphiW}&=&
\chi\,e^{{1\over k+1}\bar X}.\qquad
\EA\EE
Similarly, the bosonic screening operator~\cite{[BO]}
\BE
\Bigl(
\DbarphiW\,\betaW
+3(k+1)^2\,\BARPSI2\,\PSI1
+ (k+1)\Bigl[3\PSI1\,\BARPSI1\,\betaW
 - 2\DphiW\,\betaW
 + \d\betaW\Bigr]\,\Bigr)\,\betaW^{-2 - k}\,e^{-\barphiW-(k+1)\phiW}
\EE
evaluates under \req{fromwak} as
\BE
\Bigl(
   \DbarX - 3(k+1)\chi\,\bar\chi  -
   2(k+1)\DX\Bigr)\,e^{-\bar X-(k+1)X},
\EE
thus reproducing the third screening in the bosonized matter
theory~\req{Xmatter}.  That the Wakimoto screening operators `localize' in
the matter sector under the mapping to the new representation appears to be a
general fact (cf.~\cite{[S-sing]}), and might be used as an alternative {\it
definition\/} of the matter sector.

It should be pointed out that the
screening charges written in terms of $\chi$, $\bar\chi$, $\DbarX$ and $\DX$
imply a specific free-field realization~\req{Xmatter} of the $\N2$ matter
sector, which we only assume for the purposes of establishing a relation to
the Wakimoto representation. Using a particular free-field realization is a
matter of choice, and in that sense is not `intrinsic'$\!$, as is stressed in
the $\N2$ case by the existence of a different free-field realization of the
algebra \req{N2matter}, which we are now going to consider.

\medskip

Now, consider an alternative realization of the $\N2$
matter~\cite{[GS3],[BLNW]} in terms of a fermionic $bc$-system
$\bar\omega\,\omega$, a `Liouville' current $\DcF$ and a free boson with the
energy-momentum tensor $T_0=\half\DcX\,\DcX- (\half\AM + \AP) \d\DcX$ (with
central charge $d_0=13-6(k+1)-\frac{6}{k+1}$, hence $k\neq-1$).  This
realization reads
\BE\new\BA{rcl}
\Tm &=& T_0
  - \half\DcF\,\DcF + \AP\d\DcF
  -\frac{3}{2}\,\omega\,\d\bar\omega
  -\half\,\d\omega\,\bar\omega\,,\\
\Hm &=& \AM \DcF + \omega\, \bar\omega\,,\\
\barGm &=&
       \bar\omega\,T_0
       - \half\,\bar\omega\,\DcF\,\DcF
       +(\AP + \half\AM)\,\bar\omega\, \d \DcF
       +\half(1 - \AM^2) \d^2\,\bar\omega
       -\d\,\bar\omega\,\bar\omega\,\omega + \AM\d\,\bar\omega\,\DcF\,,\\
\Gm &=& \omega\,,\\
\multicolumn{3}{c}{
\AP=\frac{1}{\sqrt{2(k+1)}}\,,\qquad
\AM=-\frac{1}{\AP}\,.}
\EA\label{mymatter}\EE
The Wakimoto fields with the same operator products as in \req{wakope} can
now be constructed in the following way:
\BE\new\BA{rcl}
\betaW &=& e^{{2\over k}(U - F)}\,,\nonumber\\
\gammaW &=&
   \Bigl(\DF - \sqrt{\frac{k + 1}{2}}\, \DcX +
  \half \psi\, \bar\psi+ \sqrt{\frac{k + 1}{2}}\, \bar\psi\, \DcF\, \bar\omega -
   \sqrt{\frac{k + 1}{2}}\, \bar\psi\, \DcX\, \bar\omega +
       (k+1) \bar\psi\, \d\bar\omega +
       \half\omega\, \bar\omega  \Bigr)\,e^{-{2\over k}(U - F)}\,,\nonumber\\
\DphiW &=& \frac{1}{k}\,\DF - \frac{1}{k}\,\DU +
  \frac{1}{\sqrt{2(k+1)}}\,\DcF -
  \frac{1}{\sqrt{2(k+1)}}\,\DcX\,,\nonumber\\
\DbarphiW &=& -\frac{k+1}{k}\DU + \frac{k+1}{k}\DF
  - \sqrt{\frac{k+1}{2}}\,\DcF -
   \sqrt{\frac{k+1}{2}}\,\DcX + (k+1)\bar\psi\,\d\bar\omega +
   (k+1)\d\bar\psi\,\bar\omega\,,\nonumber\\
\BARPSI1 &=&
   \Bigl(\omega +
   \sqrt{\frac{k+1}{2}}\,\bar\psi\, \DcF -
   \sqrt{\frac{k + 1}{2}}\,\bar\psi\, \DcX  - (k+1)\,\d\bar\psi\Bigr)\,
e^{-{1\over k}(U - F)}\,,\nonumber\\
\PSI1 &=& -\bar\omega\,e^{{1\over k}(U - F)}\,,\nonumber\\
\BARPSI2 &=&  \bar\psi\, e^{{1\over k}(U - F)}\,,\qquad\qquad\qquad
\PSI2~{}={}~
  \Bigl(\psi
   -\sqrt{\frac{k + 1}{2}}\, \DcF\, \bar\omega +
   \sqrt{\frac{k + 1}{2}}\, \DcX\, \bar\omega  - (k+1) \d\bar\omega\Bigr)\,
e^{-{1\over k}(U - F)}\,.\nonumber
\EA\label{fromwak2}\EE
As before, these formulae provide a bridge between our realization \req{eco}
and the Wakimoto representation of $\SSL21$; however, the representation
\req{eco} (with the $\N2$ matter generators expressed as in \req{mymatter})
will be reproduced if we start with the `{\it second\/}' Wakimoto
representation of ref.~\cite{[BKT]}, which can be
written~as\,\footnote{Certain rescalings of the type of eq.~\req{Lrescale}
would slightly simplify the expressions for the Wakimoto fields, at the same
time introducing factors in front of some terms in~\req{wak2}; in the present
normalization the formulae \req{wak2} have as much as possible (e.g., the
Cartan subalgebra) in common with the `first' Wakimoto
representation~\req{wak}.}
\BE\new\BA{rcl}
E_1 &=&  \PSI2\,\betaW\ - (k+1)\,\DphiW\,\PSI1 - (k+1)\,\d\PSI1\,,\qquad
E_2~{}={}~\BARPSI2\,,\qquad E_{12}~{}={}~\betaW\,,\\
\Hplus &=&  -\half\DbarphiW + \half(k+1) \DphiW - \half\PSI1\,\BARPSI1
 + \half\PSI2\,\BARPSI2\,,\\
\Hminus &=&
    -\half\DbarphiW - \half(k+1) \DphiW - \half\PSI1\,\BARPSI1 -
      \half\PSI2\,\BARPSI2 + \gammaW\,\betaW\,,\\
F_1 &=& -\BARPSI2\,\gammaW + \BARPSI1\,,\\
F_2 &=&\PSI2\,\gammaW\,\betaW -\DbarphiW\,\PSI2 - (k+1)\,\DphiW\,\PSI1\,\gammaW
 - \PSI2\,\PSI1\,\BARPSI1 - (k+1)\,\d\PSI1\,\gammaW + k \d\PSI2\,,\\
F_{12} &=&
 -\gammaW\gammaW\betaW + \DbarphiW\gammaW + (k+1)\DphiW\gammaW
- k\d\gammaW + \PSI1\BARPSI1\gammaW - \PSI2\BARPSI1 +
   \PSI2\BARPSI2\gammaW\,
\EA\label{wak2}\EE
It is amusing to see how the various pieces of the $\N2$ matter generators
\req{mymatter} find each other when eqs.~\req{fromwak2} are inserted
into~\req{wak2}.

\medskip

Thus, allowing the $\N2$ matter generators to be bosonized through free
fields, we can relate the corresponding Fock space image of the
representation~\req{eco} to the respective Wakimoto representation.  Not
surprisingly \cite{[BLNW]}, the two different bosonizations of the $\N2$
matter are thus related to the respective inequivalent simple root systems
of~$\SSL21$.  By restricting to $\OSP12$ and $\SL2$, one can work out similar
relations for representations of these algebras.

\section{Concluding remarks}\lvm
We have shown that any non-critical $\N2$ string in the conformal gauge
reformulates as the $\SSL21$ theory tensored with two bosonic and two
fermionic `Drinfeld--Sokolov' ghost systems.  The construction of the
$\SSL21$ currents appears to be `rigid'$\!$, admitting no free parameters,
modulo the above-mentioned automorphisms of the defining operator products
(it is a useful exercise to lift the spectral flow transform
\cite{[SS],[LVW]} to the $\SSL21$ algebra).

\smallskip

The present construction may provide new group-theoretical `{\it
explanations\/}' for a number of hidden supersymmetry algebras that arise in
non-critical string theories.

\smallskip

An important problem left for the future investigation is that of a precise
comparison of the {\it physical states\/} (i.e., the BRST cohomologies) of
the $\SSL21$ theory and the $\N2$ string. The problem may be feasible by the
methods developed in conformal field theory. In the $\N2$ matter sector, we
have just the corresponding {\it Verma\/} modules, hence no need for
screening charges to project from the Fock modules. However, the two
$\beta\gamma$ systems that participate in the construction
\req{EEE}--\req{F12} are potentially more interesting from the cohomological
point of view: as we have noted, the $\SSL21$ currents depend only on the
respective $\beta\gamma$ {\it currents\/}, which excludes, by the well-known
argument (see \cite{[FMS]} and numerous subsequent applications of the idea),
the zero mode of a spin-0 fermion that participates in the bosonization of
the $\beta\gamma$ system.

\smallskip

One may wonder how special the realization \req{EEE}--\req{F12} of the
$\SSL21$ algebra is. A related question is whether one can
reconstruct the non-critical $\N2$-string fields from an `abstract' $\SSL21$
algebra {\it and\/} the Drinfeld--Sokolov ghosts (we have seen
in~\req{identity} that only with these ghosts added do the \emt s match and
thus there is a chance to recover the $\N2$ string). The $\N2$ matter sector
can indeed be recovered, in quite abstract terms, via the hamiltonian
reduction.  The two bosonic and two fermionic Drinfeld--Sokolov ghost systems
can roughly (in fact, modulo some mixing with the other free fields) be
traded for the same number of bosonic and fermionic $\N2$ {\it string\/}
ghosts. The problem is therefore concentrated in the $\N2$ Liouville sector
\req{Lopes}.  The current $\bar\psi\psi$ could then be found from the second
equation in~\req{Hminus} (assuming the matter $U(1)$ current $\Hm$ known from
the hamiltonian reduction).  This leaves us with the first equation
in~\req{Hminus}, which does not, however, allow us to determine $\Dbarphi$
and $\Dphi$ in terms of the given $\SSL21$ currents and the other ingredients
which we assumed known:  the matter $U(1)$ current $\Hm$ and the ghost
currents $B\,C$, $\BETA\,\GAMMA$, $\tBETA\,\tGAMMA$, and $\cB\,\cC$; only the
combination $-\Dbarphi + (k+3)\Dphi$ can be determined.  That finding
$\Dbarphi$ and $\Dphi$ and hence all the $\N2$ string ingredients should
require some special assumptions is also supported indirectly by the fact
that the $\N2$ string has an $\N4$ symmetry \cite{[BLLS]}; if we had been
able to identify the $\N2$ non-critical string in a general representation of
$\SSL21\tensor({\rm DS\ ghosts})$, we would have had to conclude that the
$\N4$ generators could be constructed in any such representation as well.
Anyway, it would be interesting to find the mildest requirements that would
make that possible.

\smallskip

As we have seen, the module in which the $\SSL21$ currents
\req{EEE}--\req{F12} are represented can be constructed by tensoring the
$\N2$ matter {\it Verma\/} module with free field Fock spaces. This allows us
to expect that the $\SSL21$ singular vectors would evaluate as the $\N2$
superconformal singular vectors (thus it happens in the $\SL2$ case
\cite{[GS3],[S-sing],[GrR]}). However, the effects due to the different types
of the $\N2$ \cite{[ST23]} and $\SSL21$ singular vectors and the {\it
sub\/}singular vectors of these algebras will have to be accounted for, which
appears to be an interesting problem.

\paragraph{Acknowledgements.}
I am grateful to P.~Bowcock, A.~Taormina and A.~Tseytlin for stimulating my
interest in $\N2$ strings and for very interesting discussions.  Extensive
computations of the operator product expansions have been done using the {\sc
Mathematica} package {\tt OPEdefs.m} by K.~Thielemans~\cite{[Thiel]}.  This
work was supported in part by the RFFI grant 96-02-16117 %%%%%%%% Fainberg
and by the European Community grant INTAS-93-2058.

\def\theequation{A.\arabic{equation}}
\subsection*{Appendix}\lvm
The $\SSL21$ algebra consists of four bosonic currents $E_{12}$, $F_{12}$,
$\Hminus$ and $\Hplus$, and four fermionic ones $E_1$, $E_2$, $F_1$ and
$F_2$.  The non-vanishing OPEs read,
\BE\new\BA{rclrcl}
  H^\pm(z)\, H^\pm(w) &=&  \mp{k/2\over(z-w)^2}\,,&
  E_{12}(z)\, F_{12}(w) &=&   {k\over(z-w)^2} + 2 {\Hminus\over z-w}\,,\\
  \Hminus(z)\, E_{12}(w) &=& {E_{12}\over z-w}\,,&
  \Hminus(z)\, F_{12}(w) &=& {-F_{12}\over z-w}\,,\\
E_{12}(z)\, F_1(w) &=& {E_2\over z-w}\,,&   E_{12}(z)\, F_2(w) &=& -{E_1\over
z-w}\,,\\
F_{12}(z)\, E_1(w) &=& -{F_2\over z-w}\,,&  F_{12}(z)\, E_2(w) &=& {F_1\over
z-w}\,,\\
H^\pm(z)\, E_1(w) &=& \half {E_1\over z-w}\,,&
H^\pm(z)\, F_1(w) &=& -\half {F_1\over z-w}\,,\\
H^\pm(z)\, E_2(w) &=& \mp\half {E_2\over z-w}\,,&
H^\pm(z)\, F_2(w) &=& \pm\half {F_2\over z-w}\,,\\
  E_1(z)\, F_1(w) &=& - {k\over(z-w)^2} + {\Hplus - \Hminus\over z-w}\,,&
  E_2(z)\, F_2(w) &=&  {k\over(z-w)^2} + {\Hplus + \Hminus\over z-w}\,,\\
E_1(z)\, E_2(w) &=&  {E_{12}\over z-w}\,,&
F_1(z)\, F_2(w) &=&  {F_{12}\over z-w}
\EA\label{sl21opes}\EE
all the other OPEs being regular. The Sugawara \emt\ reads
\BE
T_{\rm Sug} = \frac{1}{k+1}\Bigl(\Hminus\, \Hminus - \Hplus\,\Hplus +
\half E_{12}\, F_{12} + \half F_{12}\, E_{12} +
\half E_1\,F_1 - \half F_1\,E_1 -
\half E_2\,F_2 + \half F_2\,E_2\Bigr)
\label{N2Sug}\EE

The $\N2$ matter, which is one of the ingredients to construct the
$\SSL21$ currents, is defined by
\BE\new\BA{rclrcl}
\Tm(z)\,\Tm(w)&=&\multicolumn{3}{l}{ {{\cm/2}\over(z-w)^4} +
2 {\Tm(w)\over(z-w)^2} +
{\d \Tm(w)\over z-w}\,,}\\
\Hm(z)\,\Hm(w)&=&  {{\cm/3}\over(z-w)^2}\,,&
\Tm(z)\,\Hm(w)&=& {\Hm(w)\over(z-w)^2} + {\d \Hm(w)\over z-w}\,,\\
\Tm(z)\,\Gm(w)&=& \threehalves {\Gm(w)\over(z-w)^2} + {\d \Gm(w)\over z-w}\,,&
\Tm(z)\,\barGm(w)&=& \threehalves {\barGm(w)\over(z-w)^2} +
{\d \barGm(w)\over z-w}\,,\\
\Hm(z)\,\Gm(w)&=& {\Gm(w)\over z-w}\,,&
\Hm(z)\,\barGm(w)&=& -{\barGm(w)\over z-w}\,,\\
\barGm(z)\,\Gm(w)&=&\multicolumn{3}{l}{ {{\cm/3}\over(z-w)^3} -
{\Hm(w)\over(z-w)^2} + {\Tm(w) - {1\over2} \d \Hm(w)\over z-w}}
\EA\label{N2matter}\EE

For completeness, let us also list the non-vanishing $\OSP12$ OPEs:
\BE\new\BA{rclcrcl}
\Iplus(z) \Iminus(w)&=&{ k \over(z-w)^2} +{2\Inaught\over z-w},
&{}&\Inaught(z)\Inaught(w)&=&{k/2\over(z-w)^2},\\
 \iplus(z) \iminus(w) &=& {-2k \over(z-w)^2} - 2{\Inaught\over z-w},
&{}& \Inaught(z)\Ipm(w)&=&\pm{\Ipm\over(z-w)}\\
 \Inaught(z) \ipm(w) &=&\pm\half{\ipm\over z-w},
&{}& \Ipm(z) \imp(w) &=& -{\ipm\over z-w}, \\
\ipm(z)\ipm(w) &=&\mp2 {\Ipm\over z-w}
\EA\label{osp12OPE}\EE

\end{document}